# A Novel Mutual Insurance Model for Hedging Against Cyber Risks in Power Systems Deploying Smart Technologies

Pikkin Lau, *Student Member, IEEE*, Lingfeng Wang, *Senior Member, IEEE,* Wei Wei, Zhaoxi Liu, *Member, IEEE*, and Chee-Wooi Ten, *Senior Member, IEEE*

*Abstract*—In this paper, a novel cyber-insurance model design is proposed based on system risk evaluation with smart technology applications. The cyber insurance policy for power systems is tailored via cyber risk modeling, reliability impact analysis, and insurance premium calculation. A stochastic Epidemic Network Model is developed to evaluate the cyber risk by propagating cyberattacks among graphical vulnerabilities. Smart technologies deployed in risk modeling include smart monitoring and job thread assignment. Smart monitoring boosts the substation availability against cyberattacks with preventive and corrective measures. The job thread assignment solution reduces the execution failures by distributing the control and monitoring tasks to multiple threads. Reliability assessment is deployed to estimate load losses convertible to monetary losses. These monetary losses would be shared through a mutual insurance plan. To ensure a fair distribution of indemnity, a new Shapley mutual insurance principle is devised. Effectiveness of the proposed Shapley mutual insurance design is validated via case studies. The Shapley premium is compared with existent premium designs. It is shown that the Shapley premium has high indemnity levels closer to those of Tail Conditional Expectation premium. Meanwhile, the Shapley premium is nearly as affordable as the coalitional premium and keeps a relatively low insolvency probability.

## NOMENCLATURE

### A. Acronym

| | |
|---|---|
| TGs | Transmission Grids |
| ICTs | Information and Communications Technologies |
| IEDs | Intelligent Electronic Devices |
| CPSs | Cyber-Physical Systems |
| SCADA | Supervisory Control And Data Acquisition |
| IDS | Intrusion Detection System |
| TTF | Time-To-Failure |
| SCT | Substation Compromise Time |
| ENM | Epidemic Network Model |
| BN | Bayesian Network |
| CVSS | Common Vulnerability Scoring System |
| HMI | Human-Machine Interface |
| WAP | Wireless Access Point |
| RTUs | Remote Terminal Units |
| EMUs | Energy Management Units |
| VaR | Value at Risk |
| TCE | Tail Conditional Expectation |
| ELC | Expected Load Curtailment |
| EFC | Expected Faulty-Bus Count |
| OPF | Optimal Power Flow |
| MCS | Monte Carlo Simulation |
| SD | Standard Deviations |
| CoVs | Coefficients of Variation |
| FOR | Forced Outage Rate |
| SoI | Strengths of Interdependence |

### B. Notation

| | |
|---|---|
| $CC$ | a control center |
| $S$ | a substation |
| $T_c$ | Substation Compromise Time |
| $v_h$ | a network vulnerability |
| $t_s$ | vulnerability sojourn time |
| $J_1, J_2, J_3$ | numbers of assigned job threads |
| $Y_1, Y_2$ | durations of the task operation |
| $U$ | residual time of the job thread executing the task |
| $\lambda$ | failure rate of the substation CPS elements |
| $\mu$ | repair rate of the substation CPS elements |
| $(\lambda_b, \mu_b)$ | baseline failure and repair rates |
| $(\lambda_i, \mu_i)$ | smart-monitoring state failure and repair rates |
| $(\lambda_c, \mu_c)$ | composite smart-monitoring failure and repair rates |
| $p(v_h)$ | probability of exploiting $v_h$ |
| $\varsigma(v_h)$ | score of the vulnerability $v_h$ |
| $p(v_h \wedge c_h)$ | probability that $v_h$ is exploited by $c_h$ |
| $p(c_h\|v_h)$ | conditional probability of successfully exploiting |
| $p(c_h)$ | total probability of successful exploitation on $v_h$ |
| $\zeta$ | set of adjacent nodes of a given graphical node |
| $\overrightarrow{T}_{rec}$ | epidemic recovery time vector |
| $\overrightarrow{T}_{epi}$ | epidemic infection time vector |
| $T_{rec}$ | epidemic recovery time of the substation |
| $T_{epi}$ | epidemic infection time of the substation |
| $\hat{T}_c$ | sampled substation compromise time |
| $\hat{T}_r$ | sampled substation repair time |
| $\mathbb{B}_\zeta$ | binomial variate of the adjacent node infectivity |
| $\varepsilon$ | basic reproduction number |
| $c$ | graphical edge coupling number |
| $Z_{epi}$ | external epidemic infection time |
| $R_{epi}$ | external epidemic recovery time |
| $p_{atk}$ | probability of cyberattack infection |
| $\mathbb{P}_v$ | correlated uniform variate of state sampling |
| $\mathbf{1}_{\{\cdot\}}$ | binary indicator function |
| $\boldsymbol{S_x}$ | set of substations |
| $\boldsymbol{K_x}$ | load curtailment vector (MW) |

This work was supported by U.S. National Science Foundation (NSF) under awards ECCS1739485 and ECCS1739422.

P. Lau, L. Wang, and Z. Liu are with Department of Electrical Engineering and Computer Science, University of Wisconsin-Milwaukee, Milwaukee, WI 53211, USA.

W. Wei is with Department of Mathematical Sciences, University of Wisconsin-Milwaukee, Milwaukee, WI 53211, USA.

C.-W. Ten is with Department of Electrical and Computer Engineering, Michigan Technological University, Houghton, MI 49931, USA.

| | |
|---|---|
| $\nu$ | time step of the reliability assessment |
| $\boldsymbol{B}$ | substation susceptance vector |
| $\boldsymbol{\theta}$ | vector of the substation voltage angles (rad) |
| $\boldsymbol{G}$ | vector of the available generation (MW) |
| $\boldsymbol{G_{cap}}$ | generation capacity vector (MW) |
| $\boldsymbol{D_{cap}}$ | load capacity vector (MW) |
| $\boldsymbol{F}$ | transmission power flow vector (MW) |
| $\boldsymbol{F_{cap}}$ | thermal limit vector of the transmission lines (MW) |
| $EN(\boldsymbol{S_x})$ | enabling function of the substations |
| $\mathbf{1}_{\{\cdot\}}$ | true/false binary indicator of a conditional statement |
| $*$ | element-wise product operator |
| $\pi_1$ | TCE Premium |
| $\pi_2$ | Coalitional Premium |
| $\pi_3$ | The proposed Shapley Premium |
| $\mathbb{C}_q$ | Shapley value of TG $q$ |
| $U$ | universal set including all participating TGs |
| $\varepsilon_{q,k}$ | Shapley cost of the TG $q$ among $k$ TGs in $S$ |
| $S$ | subset of the selected TGs |
| $\delta_q$ | cumulative loss distribution |
| $y$ | number of TGs in the universal set |
| $k$ | number of TGs which submit their claims |
| $\Gamma^*_{q,k}$ | base indemnity of TG $q$ among $k$ TGs in $S$ |
| $\psi(\cdot)$ | scaling function of $\Gamma_q$ |
| $\Gamma^\psi_{q,k}$ | indemnity of TG $q$ among $k$ TGs in $S$ |
| $I^\psi_q$ | indemnity of TG $q$ |
| $\Phi(\pi)$ | probability of insolvency |

## I. INTRODUCTION

THE technological transformation from the conventional power grid to smart grid has been drastically galvanized by the rapid development and deployment of ICTs. With ever-increasing market values in the smart grid, the mass adoption of IEDs and other cyber-capable equipment in CPSs leads to increased network connectivity and higher operation uncertainty. Various security challenges in operations may sabotage or paralyze contemporary power grids [1]-[3]. For example, configuration change attack is a common cyber-threat where IEDs receive commands to manipulate the settings. Distance relays may receive malicious commands leading to unnecessary trips to transmission lines and disruption to substation operation. To mitigate the emerging cyber-threats, diverse cybersecurity technologies have been developed. For example, transient detection algorithms and obfuscation strategy can be developed for mitigating device risks [4] and enhancing cybersecurity [5]. To protect the SCADA systems, IDS is a promising solution against potential cyberattacks. An advanced machine learning based IDS with high accuracy was developed [6]. A reinforcement learning algorithm was tailored to obtain the stochastic strategies that minimize load curtailment subjected to coordinated attacks [7]. IDS involving supervised learning should have collected raw data preprocessed and labeled for the classification purpose. An ensemble approach was developed for the comparative feature extraction [8]. Security models are in place to evaluate and defend against the cyberattacks. A Cyber-Net model was developed to capture the cyber intrusion of switching attacks in a comprehensive power CPS including firewalls, SCADA and IEDs in a substation [9]. A data-driven stochastic game model was proposed to evaluate the cross-layer security of CPSs [10]. A matrix-based model was developed to capture the cyber-physical coupling behavior of IDS [11]. Strategic competition between defense and intrusion across the temporal state transition can be described by the Markov decision process [12]. Security metrics such as TTF are proposed to gauge the long-term risk of a network [13]. Investment in the enhancement of CPSs may be beneficial to reducing the interruption costs in the face of cyberattacks. Smart monitoring based on preventive and corrective measures was proposed for cyber network of the power system [14]. Job assignment strategy was designed to boost the reliability for vehicular clouds in [15].

Conventionally, technological solutions are considered the main approach to mitigate cyber risks. In this work, cyber-insurance is further devised as a promising financial instrument to hedge against the emerging cyber risks. A cyber-insurance design framework typically includes cyber risk modeling, cyberattack-induced loss estimation, and insurance premium calculation. In [16], stochastic cybersecurity insurance pricing models for graphical networks was proposed. In [17] and [18], the potential cyber-insurance models for power systems were explored. Accounting for various cyber-intrusion scenarios, the cyber-insurance designs specific to the implications of cyberthreats on power system reliability were developed. In [17], an insurance principle was developed upon the system reliability assessment at varying intrusion tolerance capabilities of the substation SCADA servers. In [18], a Bayesian Network graphical model for integrated cybersecurity-reliability was applied to a coalitional insurance design. In this work, a new mutual cyber insurance model will be developed for power grids deploying smart technologies.

The main contributions of this paper are summarized as follows:

- A novel mutual cyber-insurance framework based on the Shapley value of the cooperative game is devised. Load loss distributions are extracted from the mutual insurance participants to formulate cost values, ultimately obtaining reduced costs in the cooperative game.
- An integrated reliability evaluation model considering substations deploying smart monitoring and job thread assignment technologies is developed aiming to enhance the system robustness against cyberattacks.
- A state-sampling cyber epidemic model is integrated into the Bayesian Network cyber vulnerability model. This integrated model is devised considering the propagation of cyberattacks and network correlation.

The remainder of this paper is organized as follows. Section II introduces cybersecurity in the substation CPSs coupled with the smart technologies. In Section III, the insurance principle is elaborated. Section IV presents the case studies including numerical simulations. Section V draws conclusions for this paper.

### A. Related Work

To clarify the advancement of this work, it would be beneficial to review the most relevant work. This work continues our quest for efficient cybersecurity insurance designs, following [17] and [18]. In [17], a cyber-insurance model based on estimating the cyber risk and reliability of interdependent TGs was proposed. Reference [18] presents an alternative cyber-insurance design based on the coalitional insurance platform [19]. In this work, a new cyber-insurance framework based on Shapley cooperative game is proposed to integrate the cyber-insurance model [17] into the coalitional cyber-insurance model [18]. A prototypical epidemic spreading model [16] was used to explore the upper and lower bounds of cyberattack infection probabilities.

The proposed cyber epidemic model inspired by [16] is conceived to further estimate the long-term infectious vulnerability risk on the graphical cyber epidemic network model coupled with power systems. While the epidemic model in [16] may be effective in estimating the upper bounds of infection probabilities in various scenarios, explicit mathematical form to evaluate the physical and monetary losses incurred in CPSs by the cyberattack needs to be further developed. The proposed epidemic network model is thus specifically tailored to the reliability-based load curtailment estimation. This study examines the effectiveness of the proposed epidemic network model considering smart technologies including smart monitoring [14] and job thread assignment [15] for improving the reliability of TGs. The results of reliability assessment are further integrated in the proposed insurance premium design. Shapley value is utilized in the proposed insurance premium design in the proposed insurance premium design as a unique solution of the cooperative game which optimizes the price of anarchy [20]-[22]. The interdependence of CPSs was addressed in [23] by creating the probability table without explicitly assigning the SoI. In this work, the SoI are directly incorporated into the state sampling process which are then reflected in load loss profiles of TGs. The comparison with the related work described above is summarized in Table I.

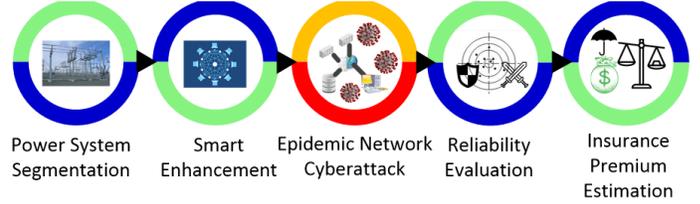

Fig. 1. The major steps in developing the proposed cybersecurity mutual insurance model.

TABLE I SUMMARY OF COMPARISON WITH RELATED WORK

| | This work | [18] | [17] | [19] | [14] | [15] | [16] | [23] |
|---|---|---|---|---|---|---|---|---|
| RA | ✓ | ✓ | ✓ | | ✓ | | | |
| CPE | ✓ | ✓ | ✓ | | | | ✓ | |
| SM | ✓ | | | ✓ | ✓ | | | |
| JA | ✓ | | | | | ✓ | | |
| SCG | ✓ | | | | | | | |
| MI | ✓ | ✓ | | ✓ | | | | |
| LI | ✓ | ✓ | ✓ | | | | | ✓ |
| CSM | ✓ | ✓ | ✓ | | | ✓ | | |
| GCVN | ✓ | ✓ | | | | | | |
| CEM | ✓ | | | | | | ✓ | |

\* RA = Reliability Assessment, CPE = Cyber Premium Evaluation, SM = Smart Monitoring, JA = Job Assignment, SCG = Shapley Cooperative Game, MI = Mutual Insurance, LI = Loss Interdependence, CSM = Cyber-Security Metric, GCVN = Graphical Cyber Vulnerability Network, CEM = Cyber Epidemic Model.

## II. PROPOSED EPIDEMIC CYBER-PHYSICAL SYSTEM MODEL

A goal of this study is to gauge the risk of cyberattacks on the individual TGs to determine economical insurance pricing strategies. Fig. 1 conveys the proposed mutual insurance framework as multiple steps: (a) The power system configuration under study should be segmented according to the TGs ownership. (b) Within respective TG substations, smart monitoring and server job assignment are enforced to enhance the substation reliability subject to cyberattacks. (c) Accounting for the cyber connection across the TGs, an ENM is established to stochastically evaluate the long-term impact of cyberattacks. (d) Reliability-based optimal power flow is conducted to estimate the load loss profiles of respective TGs. (e) The insurance premium of each TG is computed based on the corresponding marginal distribution of the loss.

### A. Epidemic Network Model

Fig. 2 illustrates the attack graph of the proposed ENM. The vulnerability $v_h$ is denoted as a colored oval VUL. Each node represents a vulnerability. In the proposed ENM, two types of anomalies, ROB and DoS, are considered. ROB attack decrypts the control center's server privilege by iterating queries to a control server. DoS attack on the substation server is triggered by unauthenticated clients issuing specially crafted messages. The successful exploitation condition of vulnerability $v_h$ occurs

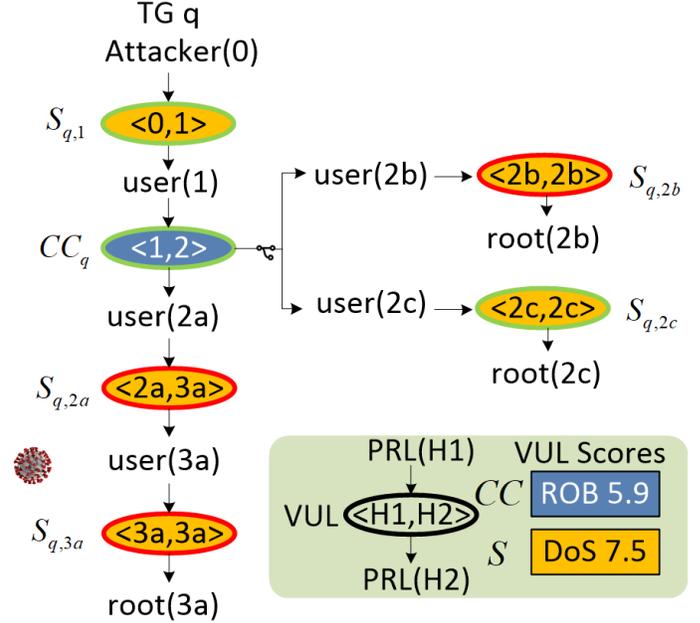

Fig. 2. Attack graph of the proposed Epidemic Network Model.

when the server privilege is obtained by the attacker, denoted as $c_h$.

Vulnerability scores are determined by CVSS comprising the base score, temporal score, and environmental score that take a wide range of attack-related factors into account, including confidentiality, integrity, availability, attack complexity, privileges required, and exploit code maturity [24].

In Fig. 2, the attacker may compromise the substation $S_{q,1}$ to start the attack on the $q$-th TG. Specifically, the attacker deploys anomaly DoS(1) to gain access to the server privilege user(1) of $S_{q,1}$ by exploiting <0, 1>. Once $S_{q,1}$ is compromised, adjacent <1, 2> of the control center $CC_q$ can be exploited in a similar manner. Vulnerabilities in cascade are exploited sequentially. In $TG_q$, the substations and the control center are laid out according to cyber connections in the attack graph. Power dispatching action is feasible along the good routes connecting healthy nodes. Then the good routes are obtained using a routing algorithm such as Depth-First Search [25]. The substations outside of the good routes indicate disconnection from power generation capacity, resulting in load curtailment in the TG. More details can be referred to [26] for cyber network modeling.

Physically, exploiting the vulnerability means the attacker breaches the server firewall to gain the server privilege to manipulatively command the substation. According to the attack graph, all preceding and current vulnerabilities should be exploited to compromise a substation server. However, since the substations located at any point of the attack graph may be compromised, an external infection term is established in the Cyber Epidemic Model to include such a possibility. After the substation server is compromised, the attacker may send counterfeit commands to the protective relays to disconnect the major substation IED from grid operations.

The reliability-driven approach adopted in this study is different from contingency analysis on cascading failures. A graphical S-k contingency analysis based on extended enumeration considers the cascading failure by gradually removing the overloaded lines [27]. Worst cases with divergent load flow results are recorded to estimate the substation risk indices. Differently, in the MCS based reliability evaluation procedure, each component status is determined by comparing the random number generated and the FOR of that component. Then, a reliability-based OPF is performed for the sampled state to decide if there is load loss after re-dispatching the power to minimize the overall load loss based on the current system state. Finally, the overall reliability indices can be calculated by sampling enough system states with varying failure scenarios.

**Definition 1:** Substation Compromise Time

$$T_c = \frac{\sum_{v_h \in V} t_s(v_h) p(v_h \wedge c_h)}{p(c_h)} \tag{1A}$$

**Job Assignment:**

$$t_s(v_h) = \begin{cases} t_s(J_1, \lambda, \mu) = \frac{1}{\lambda} \\ t_s(J_2, \lambda, \mu) = \frac{1}{\lambda} + \frac{1}{2\lambda(1-p_1)} \\ t_s(J_3, \lambda, \mu) = \frac{1}{\lambda} + \frac{1}{2\lambda(1-p_1)} + \frac{1}{3\lambda(1-p_1)(1-p_2)} \end{cases} \tag{1B}$$

$$\begin{cases} p_1 = \Pr[Y_1 > U] = \frac{\mu}{\mu + \lambda} \\ p_2 = \Pr[\min\{Y_1, Y_2\} > U] = \frac{\mu}{\mu + 2\lambda} \end{cases} \tag{1C}$$

**Smart Monitoring:**

$$(\lambda_c, \mu_c) = (\sum_{i=0}^{N} \lambda_i, \frac{\lambda_c * P_{Up}}{1 - P_{Up}}) \tag{1D}$$

$$\begin{cases} P_{Up_b} = \frac{\mu_b}{\mu_b + \lambda_b} < P_{Up} = \frac{\mu_c}{\mu_c + \lambda_c} \\ \mu_b = \mu_0 \\ \mu_b \leq \mu_i, 1 \leq i \leq N \\ \lambda_b = \sum_{i=0}^{N+M} \lambda_i \end{cases} \tag{1E}$$

**Epidemic Network Model:**

$$p(v_h) = \frac{\varsigma(v_h)}{10} \tag{1F}$$

$$p(c_h | v_h) = \begin{cases} U(0.8, 1) * \mathbf{1}_{\{c_h = T\}}, h = 1 \\ p(c_h | v_h \wedge (v_1 \vee \ldots \vee v_{h-1})), n \geq h \geq 2 \end{cases} \tag{1G}$$

$$p(v_h \wedge c_h) = p(v_h) * p(c_h | v_h) \tag{1H}$$

$$p(c_h) = \sum_{l=1}^{n} p(v_l \wedge c_h) \tag{1I}$$

A security metric used extensively in reliability assessment is the SCT $T_c$. In *Definition 1*, $T_c$ quantifies the time taken by the attacker to bring the substation down. Considering the attack graph, $T_c$ is logically synthesized with a Bayesian Network [13]. The SCT $T_c$ is synthesized based on the individual sojourn times $t_s$ of the vulnerabilities $v_h$. $t_s$ is positively correlated to $T_c$ and the overall system reliability. The smart technologies considered in this work including smart monitoring [14] and job thread assignment [15]. In the job assignment, $t_s$ corresponds to the memory thread resource in the cyber-physical operational task. In smart monitoring, the CPS elements deploy preventive and corrective measures to boost substation security. The probability of exploiting vulnerability $v_h$ depends on its score $\varsigma(v_h)$. The conditional probability $p(c_h | v_h)$ can be determined by the vulnerability $v_h$ and all the preceding vulnerabilities. The total probability of successful exploitation $p(v_h \wedge c_h)$ is the summation of $p(v_h \wedge c_h)$.

### B. Cyber Epidemic Across the Transmission Grids

**Definition 2:** The proposed Cyber Epidemic Model

$$EN(S_x) = \mathbf{1}_{\{P_v \geq p_{atk}\}} \tag{2A}$$

$$p_{atk} = \frac{T_{rec}}{T_{epi} + T_{rec}} \tag{2B}$$

$$\mathbb{P}_v \sim U(0,1) \tag{2C}$$

$$\zeta = \{1, \ldots, \gamma\} \tag{2D}$$

$$\mathbb{B}_g \sim Bin(1, c) \tag{2E}$$

$$\vec{T}_{epi} = [\hat{T}_{c,1} \quad \ldots \quad \hat{T}_{c,\gamma} \quad Z_{epi}] \tag{2F}$$

$$\vec{T}_{rec} = [\hat{T}_{r,1} \quad \ldots \quad \hat{T}_{r,\gamma} \quad R_{epi}] \tag{2G}$$

$$\hat{T}_r = \{\varepsilon \sum_{\Gamma} \mathbb{B}_{\Gamma}\} \tag{2H}$$

$$T_{rec} = \max(\vec{T}_{rec}) \tag{2I}$$

$$T_{epi} = E[\vec{T}_{epi}] \tag{2J}$$

The cyber epidemic model is initiated by a malicious attack described in *Definition 2*. The cyber epidemic model that infects a vulnerability node may stochastically spread to an adjacent vulnerability node set $\zeta$. State sequence of a specific substation is determined by the infection time vector $\vec{T}_{epi}$ and recovery time vector $\vec{T}_{rec}$ based on the SCTs of $\zeta$ and binomially distributed recovery times of $\zeta$. To consider the cyber risks spreading in the large-scale network, external epidemic infection time $Z_{epi}$ and recovery time $R_{epi}$ are respectively included in augmented $\vec{T}_{epi}$ and $\vec{T}_{rec}$. The intensity of the epidemic attack can be adjusted by the basic reproduction number $\varepsilon$ and graphical edge coupling number $c$. The substation infection time $T_{epi}$ and recovery time $T_{rec}$ are estimated by the maximum and expected values of the respective vectors. The probability of cyberattack infection $p_{atk}$ is calculated using $T_{epi}$ and $T_{rec}$. Then $p_{atk}$ is compared with a uniform variate to determine whether the substation server is compromised by the cyberattack. The proposed Epidemic Network Model concerns a software-based cyberattack whose impacts would be reflected in the physical power system. When the substations become infected, their operations are compromised. Specifically, the physical loss due to the cyberattack is measured by load curtailment in the reliability analysis. The economic implication of the load loss on TGs is then evaluated in the insurance premium design.

**Optimization 1:** Reliability-based Load Curtailment Estimation

$$\min \{\sum_x K_x\}_v \tag{3A}$$

**Subject to:**

$$B\theta + G + K_x = D_{cap} \tag{3B}$$

$$|F| \leq F_{cap} \tag{3C}$$

$$0 \leq K_x \leq D_{cap} \tag{3D}$$

$$0 \leq G \leq EN(S_x) * G_{cap} \tag{3E}$$

$$EN(S_x) = \mathbf{1}_{\{P_v \geq p_{atk}\}} \tag{3F}$$

The substation state sequence $EN(S_x)$ is sampled subject to cyber epidemic described in *Definition 2*, with binary values 1 and 0 indicating the generation capacity $G_{cap}$ connected to the specific substations to be either available or offline. If the substation server is infected by the cyberattack, the attacker could breach the server root privilege and send false tripping commands to the substation relays that cause generation offline. In *Optimization 1*, $EN(S_x) * G_{cap}$ determines the upper bounds of online capacity $G$ at each time step $v$. Together with the load capacity $D_{cap}$ and thermal limit constraints $F_{cap}$, the aggregate substation load loss $\sum_x K_x$ is minimized at each time step $v$. The energy balance between the online generation supply and online load demand should always be maintained with load curtailment $K_x$ being further bounded by the load capacity $D_{cap}$.

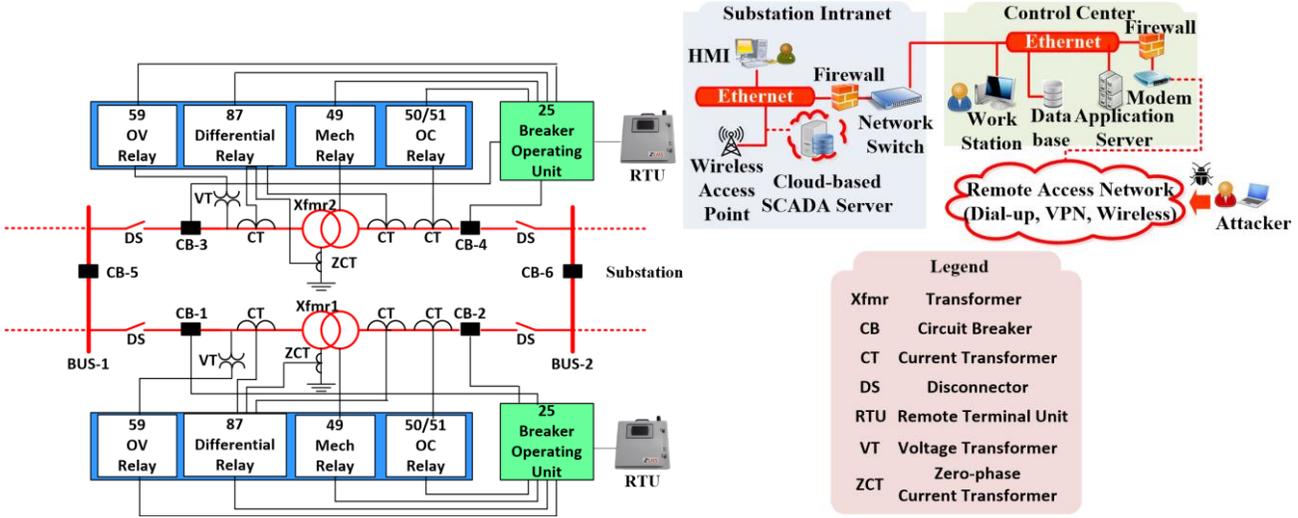

Fig. 3. A typical cyber epidemic on a substation.

Fig. 3 depicts a typical process of epidemic propagation on a substation. The malicious attacker infiltrates the firewall of the control center through remote access network connected via a modem. If the Ethernet in the control center is breached, the attacker can gain access to the data storage, application server and operation of the workstation. Since WAP controls the substation operation via RTUs, the attacker may directly compromise the substation through the WAP without going through control center. Through infecting the network switch with malware, the attacker may further compromise the substation Intranet. That is, the attacker obtains the privilege of the SCADA server, Human-Machine Interface and the WAP. If the WAP is hacked, false commands can be sent to RTUs to modify the trip settings in different relays. The breaker operating units connected to RTUs coordinate the relays to provide overcurrent protection, overvoltage protection and differential protection. By intentionally reducing the threshold value of the overcurrent relay, the circuit breakers can be falsely tripped when no physical fault condition is presented. A detailed survey further analyzed the impacts of various cyberattack scenarios in the power systems [28].

In the following subsection, the cyber-physical enhancement strategies on the substations will be presented.

### C. Substation Cyber-Physical Enhancement

To enhance power system reliability, substation-oriented smart monitoring including SCADA systems and EMU may be worth investments. To highlight the merit of the cyber-physical smart grid with sensing and remedial equipment, reliability modeling of the smart monitoring devices performed in Fig. 4(a) shows a base case of two-state reliability model with failure rate and repair rate $(\lambda_b, \mu_b)$. In Fig. 4(a), the smart monitoring reliability model has M+1 up states ($Up_0 \sim Up_M$) and N+1 ($Dn_0 \sim Dn_N$) down states, with failure rate and repair rates $(\lambda_i, \mu_i)$ among respective states. The smart monitoring model can be reduced to an equivalent composite two-state model with the composite failure rate and repair rate $(\lambda_c, \mu_c)$ [14].

For the substation servers, it is crucial to ensure IEDs within the substations with computing capability function normally. In typical operations of computing systems, portions of the memory are dynamically allocated to process-specific tasks. The scheduling strategies in [15] can be adapted to enable improved job thread assignment for our problem. Multiple server threads are scheduled to carry out the same task command of IEDs to heighten the computing dependability against uncertainties.

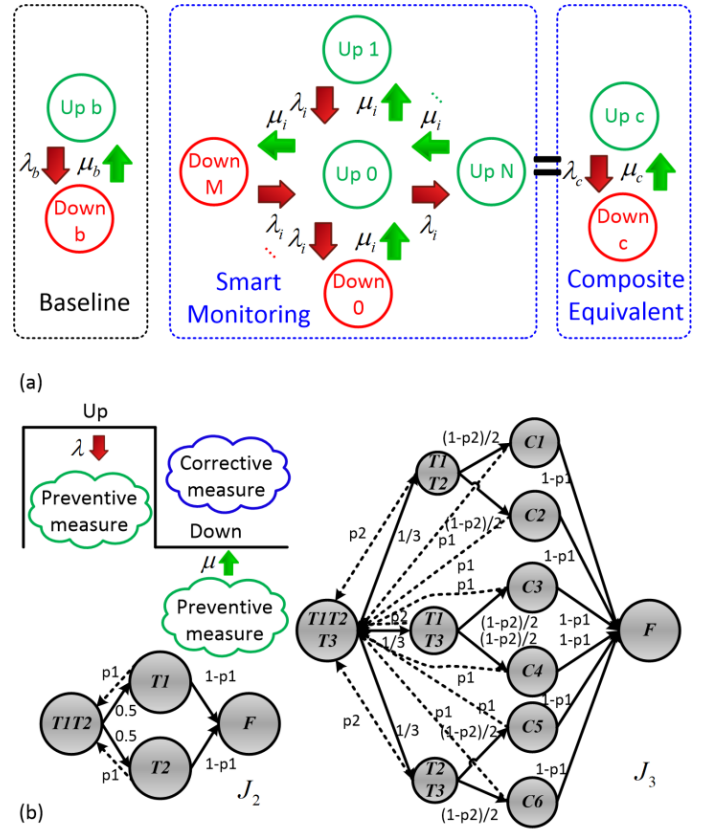

Fig. 4. (a) A baseline Markovian model vs smart-monitoring Markovian model and its composite equivalent [14]. (b) Markovian models for server job thread assignment: 2 threads ($J_2$) vs 3 threads ($J_3$) [15].

Fig. 4(b) shows a basic 2-thread ($J_2$) fault-tolerant job thread assignment procedure assigned with a critical server task in the substation SCADA server. The procedure includes total 4 states: both threads $T1T2$ carrying out single task, either thread ($T1$, $T2$) executing the same task, and the task is terminated when both threads fail $F$ to perform the task. Similarly, Fig. 4(b) also shows a 3-thread ($J_3$) fault-tolerant job thread assignment procedure with 11 states: all 3 threads $T1T2T3$ conducting single task, 2 of the threads (*T1T2, T1T3, T2T3*) carrying out the same task, within 2 selected threads one of the threads further fails (*C1~C6*), and the task is terminated when all 3 threads fail $F$ to perform the task. The sojourn time of $J_2$ and $J_3$ are $t_s(J_2, \mu, \lambda)$ and $t_s(J_3, \mu, \lambda)$, determined by the probabilities of state transition, duration of task operation, expected thread recruitment rate $\mu$ and thread residence rate $\lambda$. In this study, duration of the task operation and residual time of the job thread executing the task are assumed to be exponentially distributed for simplicity.

It will be shown in the case studies the combined application of job thread assignment and smart monitoring can achieve improved grid reliability.

### D. Strength of Interdependence

The SCT $T_c$ is the hypothetical effort where the individual substation privilege access would be obtained by the malicious attacker. Typically, SCTs of the target substations are considered mutually independent in reliability assessment. In this study, all TGs are assumed participants of the proposed mutual insurance to study SoI across the TGs. The sampled SCT vector $\tilde{T}_c$ incorporates a standard uniform variate set $\mathcal{U}$ into the SCT vector $T_c$ to produce the correlated loss pattern.

$$\tilde{T}_c = T_c * \mathcal{U} \qquad (4)$$

Indirect approach is necessary to embed the correlation factor into the uniform variate. Multivariate normal variate $N_c \sim N(\mathbf{0}, \mathbf{\Sigma})$ is handy to allow specification of the correlation $r$ in the covariance matrix $\mathbf{\Sigma}$:

$$\mathbf{\Sigma} = (1 + r)\mathcal{J}_y - I_y \qquad (5)$$

where $y$ is the number of TGs, $\mathcal{J}_y$ is the all-one matrix, and $I_y$ is the identity matrix.

Substituting $N_c = \{N_{c1}, \ldots, N_{cy}\}$ into the cumulative distribution function of the standard normal distribution $\Phi$, a set of uniform variates can be obtained:

$$\mathcal{U} = \Phi(N_c) \qquad (6)$$

where $\mathcal{U} = \{\mathcal{U}_{c1}, \ldots, \mathcal{U}_{cy}\}$ is the copula of the uniform distribution with correlation coefficient $r$.

In the next section, a cyber-insurance principle for estimating the premiums of individual TGs will be introduced.

## III. PROPOSED INSURANCE PREMIUM PRINCIPLE

### A. Fundamentals

Due to the growing adoption of ICTs in power systems, financial tools to hedge against the unforeseeable cyber-related monetary losses are emerging as an alternative or supplemental solution more recently.

A crucial characteristic of the mutual insurance is to account for the financial impacts on economically related entities. Due to the high unpredictability of cyberattack-caused losses, power system application of the mutual insurance can be especially challenging. The intended mutual insurance premium design is tailored to TGs with a relatively small insured pool and large fluctuations in indemnities. An overview on the basics and existent work is provided before getting into the detailed insurance design.

**Definition 3**: Tail Risk Measures for the loss $\mathcal{L}$
$$VaR_\varpi(\mathcal{L}) = \inf\{\ell: P(\mathcal{L} > \ell) \le \varpi\}, \ \varpi \in (0,1) \quad (7A)$$
$$\pi_1(\mathcal{L}) = TCE_\varpi(\mathcal{L}) = E[\mathcal{L}|\mathcal{L} > VaR_\varpi(\mathcal{L})] \quad (7B)$$
$$\Pr[\mathcal{L} > VaR_\varpi(\mathcal{L})] = \varpi \quad (7C)$$
$$TCE_\varpi(\mathcal{L}) > VaR_\varpi(\mathcal{L}), \ \forall \mathcal{L} \quad (7D)$$
$$\Pr[\mathcal{L} > TCE_\varpi(\mathcal{L})] \le \varpi \quad (7E)$$

In *Definition 3*, VaR and TCE are statistical indices specifically for gauging risk percentile $\varpi$. VaR is the $100\varpi\%$ percentile of the loss $\mathcal{L}$. TCE is the average of the worst $100\varpi\%$ scenarios of the loss $\mathcal{L}$. Given the same level of $\varpi$, TCE is always larger than VaR. The relations among VaR, TCE and the loss $\mathcal{L}$ are described in (7).

TCE premium design $\pi_1$ [17] is a mutual insurance allocated from the insured TGs. $\pi_1$ can gauge risk conservatively based on individual contributions to $TCE_\varpi(\sum_q \mathcal{L}_q)$. In extremely catastrophic events, $\pi_1$ would be beneficial. When the tail risk is small, $\pi_1$ may induce heavy financial burden on the TGs if no major loss events occur.

$\pi_1$ is devised with the third-party insurer operation in mind. When undesirably high premium quotes from $\pi_1$ occur, an insurance coalition among the TGs comes into play handily. The coalitional insurance manages to scale down the premium risk loading by evenly distributing the premiums across participating entities. The coalitional premium $\pi_2$ [18] is a mutual insurance based on the crowdfunding model distributing the risk affordably. $\pi_2$ offers small risk loading at the cost of small loss coverage. $\pi_2$ accounts for the fairness across the TGs. The commitment and the claim of $\pi_2$ can be flexibly set on the participants' discretion; say, the TCE premium and the expected loss. In the following subsection, a novel Shapley premium design $\pi_3$ is proposed as a middle ground between $\pi_1$ and $\pi_2$.

### B. The Proposed Shapley Premium

The Shapley value [20]-[22] was introduced as a unique set of values fairly distributed across players in the cooperative games. Several basic properties should be mentioned before the premium design is presented. In a cooperative game $G = (U, \varepsilon)$ that contains a finite player universal set $U$ whose respective costs correspond to a subset $S$ are $\varepsilon(S)$, the Shapley value of the TG $q$ is defined as follows:

$$\mathbb{C}_q(U, \varepsilon) = \frac{\sum_{S \subseteq U \setminus \{q\}} |S|!(|U|-|S|-1)![\varepsilon(S \cup \{q\})-\varepsilon(S)]}{|U|!} \quad (8)$$

Here a cooperative-game based Shapley value design is proposed for the power system cyber-insurance to achieve fair risk loading. The respective losses more evenly distributed in the proposed premium design than those in the coalitional insurance.

**Definition 4**: The proposed Shapley mutual insurance principle
$$\pi_3(\mathcal{L}_q) = \mathbb{C}_q(U, \varepsilon_{q,k}) \quad (9A)$$
$$\varepsilon_{q,k}(\mathcal{L}_q) = C_k^y \delta_q^k (1 - \delta_q^k)^{y-k} \sum_{q \in S} VaR_\varpi(\mathcal{L}_q) \quad (9B)$$
$$\Gamma_{q,k}^* = \frac{y-k}{y-1} TCE_\varpi(\mathcal{L}_q) + \frac{k-1}{y-1} \sum_{q \in U} TCE_\varpi(\mathcal{L}_q) \quad (9C)$$
$$\Gamma_{q,k}^\psi = \psi(\Gamma_{q,k}^*) = \begin{cases} \Gamma_{q,k}^*, if \sum_{q \in S} \Gamma_{q,k}^* \le \sum_{q \in U \setminus S} \mathbb{C}_q \\ \frac{\sum_{q \in U \setminus S} \mathbb{C}_q}{\sum_{q \in S} \Gamma_{q,k}^*} \Gamma_{q,k}^*, else \end{cases} \quad (9D)$$

Shapley value $\mathbb{C}_q(U, \varepsilon_{q,k})$ of the loss $\mathcal{L}_q$ serves as the Shapley premium $\pi_3$, where the universal set $U$ includes all TGs in study. Given the subset $S$ including the selected TGs, Shapley cost of the q-th TG when $k$ TG(s) submit the claim is denoted as $\varepsilon_{q,k}(S)$. The Shapley cost $\varepsilon_{q,k}(S)$ handles typical risk lower than the tail risk when the cumulative loss distributions $\delta_q$ are smaller than $VaR_\varpi(\mathcal{L}_q), q \in S$. Since the typical risk in each TG varies with $k$, the probability that the specific TGs are included in a subset $S$ is determined by an unfair coin-tossing model in $\delta_q$. The cooperative game $G$ determines each $\mathbb{C}_q(U, \varepsilon_{q,k})$ by assigning the expected values of its marginal contribution. The constraint of rationality ensures $\mathbb{C}_q(U, \varepsilon_{q,k})$ that no feasible cooperation can be formed if the cooperative cost exceeds the sum of the respective costs. In other words, the Shapley cooperative game $G$ guarantees

**Algorithm 1:** Mutual Insurance for Power System Reliability

*Input:* $J_i$, $\mu$, $\lambda$, $C_q$, $S_{q,x}$, $v_h$, $r$

*Output:* $\pi(\mathcal{L}_q)$, $\rho(\mathcal{L}_q)$, $\Gamma_q^\psi$

/*Stochastic model preparation*/
1: FOREACH TG $q$
2:    FOREACH substation $x$
3:        Evaluate $(\lambda_c, \mu_c)$ based on Smart Monitoring.
4:        Substitute$(\lambda_c, \mu_c)$ into $t_s(J_i)$ of Job Assignment.
5:    END
6:    Collect cyber network information $(C_q, S_{q,x}, v_h)$.
7:    FOREACH substation $x$
8:        By *Definition 1*,
9:            Synthesize $t_s(J_i, \lambda, \mu)$ and $v_h$ in BN.
10:           Compute $T_c$ graphically using $(C_q, S_{q,x}, t_s)$.
11:   END
12: END
/*Incorporate correlation in sampling*/
13: Synthesize $\tilde{T}_c$ via the indirect approach (4) - (6).
14: Generate $EN(S_x)$ by *Definition 2*.
15: Perform *Optimization 1* to obtain $\mathcal{L}_q$ in each TG $q$.
/*Premium estimation in Section III*/
16: Estimate premium designs $\pi_1(\mathcal{L}_q)$, $\pi_2(\mathcal{L}_q)$, $\pi_3(\mathcal{L}_q)$.
17: Estimate TG indemnities $\Gamma_q^\psi(\pi_1)$, $\Gamma_q^\psi(\pi_2)$, $\Gamma_q^\psi(\pi_3)$.
18: Procedure END

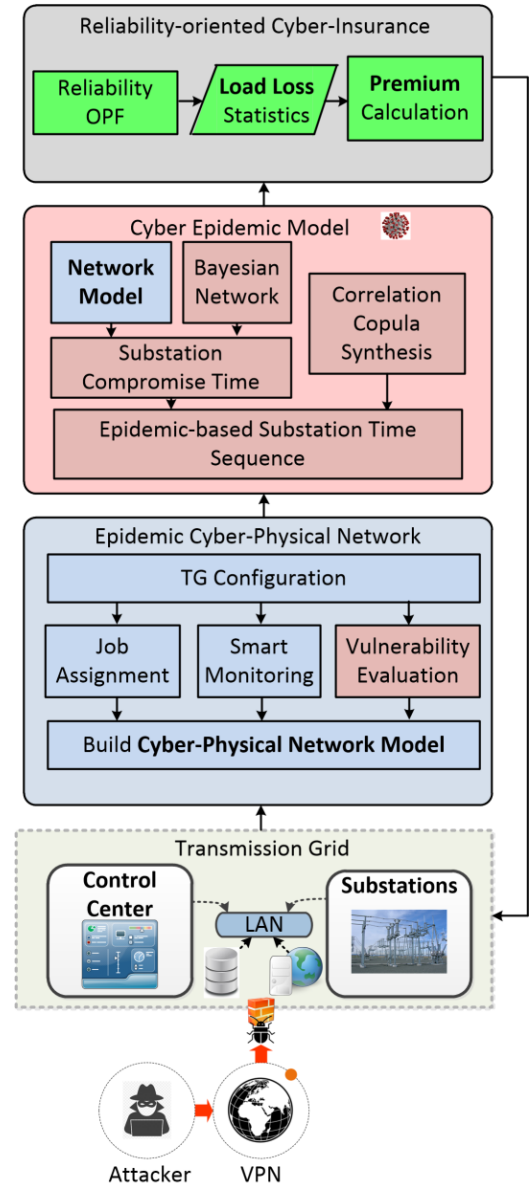

Fig. 5. Flowchart of the proposed cybersecurity mutual insurance model, comprising (I) epidemic cyber physical system modeling, and (II) cyber-insurance design.

the mutually insured individual a lower cost than its own cost. In this way, $\varepsilon_{q,k}(S)$ ensures that the Shapley premium $\pi_3(\mathcal{L}_q)$ is fairly allocated according to the loss $\mathcal{L}_q$ of the TG.

The base indemnity $\Gamma_{q,k}^{**}$ is the amount that each of the TGs can redeem from insurance when suffering from the loss event. $\Gamma_{q,k}^*$ is proportionally allocated between the self-indemnity term $TCE_\varpi(\mathcal{L}_q)$ and the group-indemnity term $\sum_{q \in U} TCE_\varpi(\mathcal{L}_q)$ summed across all the participating TGs. The group-indemnity term weighs heavily as $k$ increases, and vice versa. The scaling function $\psi(\cdot)$ ensures the budget sufficiency at various $k$ by scaling down $\Gamma_{q,k}^*$ beyond the premium $\mathbb{C}_q$. Denote the indemnity at $k$ as $\Gamma_{q,k}^\psi = \psi(\Gamma_{q,k}^*)$. The indemnity that the TG $q$ can at most redeem from a loss would be $\Gamma_q^\psi = \max_k \Gamma_{q,k}^\psi$. Like $\pi_1$ and $\pi_2$, the formulation of $\pi_3$ also incentivizes the security investment by reducing the premium payment. Besides, $\pi_3$ is a mutual insurance that intends to be a financial mutual trust. Most TGs with positive risk loading provide some margin to cushion against uncertainty. In the mutual insurance, outliers struck by unexpectedly high damages would result in negative risk loading. Losses of other TGs could partially be covered by the mutual insurance premium.

A major design goal of the insurance premium is to mitigate the risk insolvency by restraining the risk higher than the indemnity. TCE premium $\pi_1$ offers good mitigation on the risk insolvency and serves as the claim term in $\pi_3$. The nature of mutual insurance guarantees $\pi_3$ premium package is nearly as affordable as $\pi_2$. Combining the advantages of $\pi_1$ and $\pi_2$, the proposed $\pi_3$ can substantially restrain the insolvency comparable to $\pi_1$. The mutual insurance premium estimation procedure is summarized in **Algorithm 1**.

The proposed cybersecurity mutual insurance model shown in Fig. 5 can be elaborated as follows: **(1) Epidemic cyber-physical system model introduced in Section II.** The cyber attacker injects the epidemic virus through Internet that penetrates the firewall of a TG.

Within the TG, a control center and substations interconnected via the Local Area Network (LAN) are stochastically infected by the cyber epidemic. The proposed cyber-physical network model (*Definition 1*) accounts for the defensive capability of the TG via the hardware investment, software strategy development and its intrinsic vulnerabilities. With the above information, the substation state sequence (*Definition 2*) can be synthesized considering the SoI across the TGs.

**(2) Cyber-insurance design introduced in Section III.** Taking the state sequence generated by the cyber epidemic, load curtailment of the respective TGs is calculated with the reliability analysis (*Optimization 1*). Using the marginal distribution of load loss statistics, the proposed Shapley premium of the individual TGs are estimated. In the following section, the proposed Shapley premium design at various SoI

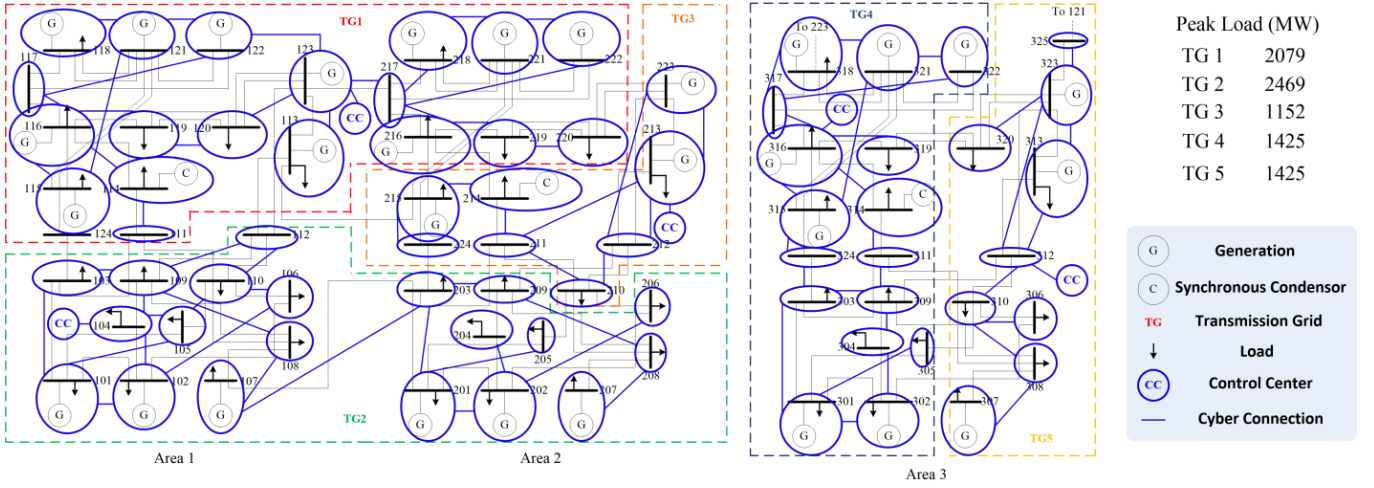

Fig. 6. TG Zones in the modified IEEE RTS-GMLC [29] including the epidemic cyber network.

| | Peak Load (MW) |
|---|---|
| TG 1 | 2079 |
| TG 2 | 2469 |
| TG 3 | 1152 |
| TG 4 | 1425 |
| TG 5 | 1425 |

and cyber-physical defense investment will be verified in the simulated case studies.

## IV. SIMULATION RESULTS

Case studies are performed to validate the proposed reliability assessment framework and cyber insurance model. As shown in Fig. 6, a benchmark IEEE RTS-GMLC is deployed [29]. The IEEE RTS-GMLC incorporates the increasing share of renewable energy resources such as wind and solar energies. To study the effectiveness of mutual insurance, the 3-area test system are divided into 5 TGs. The IEEE RTS-GMLC is further augmented by incorporating the epidemic cyberattack model. The cyberattack parameters of the epidemic network are assigned as follows: $Z_{epi} = 2000$ hrs, $R_{epi} = 4$ hrs, $\varepsilon = 2$, and $c = 0.8$.

A preliminary comparison is made on the system risk in the test system under various scenarios. Risk indices estimating load curtailment and fault coverage are adopted from [30]. Denote $LC$ as the load curtailment and $FC$ as the count of faulty buses at the $m$-th time step. The Expected $LC$ and $FC$ are defined as follows:

$$ELC = \frac{1}{N_m} \sum_{m=1}^{N_m} LC_m^T \qquad (10)$$

$$EFC = \frac{1}{N_m} \sum_{m=1}^{N_m} FC_m^T \qquad (11)$$

Parameters of the cyber-physical elements installed in the substations are listed in Table II. When the substation's smart monitoring is functional, the server is connected to other elements. Otherwise, the server is disconnected from other elements. Six scenarios are studied to demonstrate the effectiveness of the job assignment and smart monitoring. As shown in Table III, the deployment of job assignment and smart monitoring technologies effectively reduces the ELC and EFC. Reduced ELC and EFC indicate enhanced security and reliability of power supply. The job assignment facilitates Scenario 2 with 20% improvement from Scenario 1 in both ELC and EFC. With the smart monitoring technology enforced, Scenario 4 improves 7% on ELC and EFC over Scenario 1. In Scenarios 5 and 6, smart monitoring plus the job assignment can further improve several percent from Scenarios 2 and 3 with job assignment alone.

TABLE II CYBER-PHYSICAL ELEMENT PARAMETERS

| Element | Failure rate $\lambda(h^{-1})$ | Repair rate $\mu(h^{-1})$ | Reliability | State |
|---|---|---|---|---|
| Server (Attacked) | 1/9200 | 1/48 | 0.9948 | $Dn_b$ |
| Server | 1/14000 | 1/48 | 0.9966 | $Dn_0$ |
| Bus | 1/876000 | 1/6 | 0.999993 | $Dn_1$ |
| Switch | 1/45000 | 1/48 | 0.9989 | $Dn_2$ |
| Optical fiber | 1/500000 | 1/12 | 0.999976 | $Up_1$ |
| EMU | 1/877600 | 1/24 | 0.9997 | $Up_2$ |

TABLE III RELIABILITY-ASSESSMENT RESULTS OF EXAMPLE SCENARIOS

| Scenarios | | ELC (p.u.) | Improvement | EFC | Improvement |
|---|---|---|---|---|---|
| 1 | $(J_1, \mu_b, \lambda_b)$ | 0.2398 | -- | 12.915 | -- |
| 2 | $(J_2, \mu_b, \lambda_b)$ | 0.1852 | 22.77% | 9.9233 | 23.17% |
| 3 | $(J_3, \mu_b, \lambda_b)$ | 0.1471 | 38.66% | 7.7812 | 39.75% |
| 4 | $(J_1, \mu_c, \lambda_c)$ | 0.2212 | 7.66% | 11.930 | 7.63% |
| 5 | $(J_2, \mu_c, \lambda_c)$ | 0.1693 | 29.40% | 9.0138 | 30.21% |
| 6 | $(J_3, \mu_c, \lambda_c)$ | 0.1334 | 44.37% | 7.0239 | 45.62% |

The reliability-based OPF is carried out in MCS based on the state sampling method. The sampled period is 40 years with hourly time steps. The server smart technology deployment within the substations determines the SCT. Cyberattacks that penetrate the substation servers may disturb the grid operation by sending spurious commands to disconnect generation from the grid, causing physical load losses. The load loss statistics is then converted into the monetary reliability worth to estimate the cybersecurity insurance premiums.

To highlight the merits of the proposed Shapley premium design, two case groups are created to compare job thread assignment, smart monitoring, and correlation coefficients at varying degrees.

*Case Group 1:* Based on Scenario 1 $(J_1, \mu_b, \lambda_b)$ where in the substation only a single job thread is available without smart monitoring.

*Case Group 2:* Based on Scenario 6 $(J_3, \mu_c, \lambda_c)$ where the strongest job assignment and substation smart monitoring are enforced.

To explore the loss characteristics in Case Group 1, Table IV summarizes the expected values, Standard Deviations, and Coefficients of Variation under various strengths of correlation $r$. CoV is obtained from the SD being divided by the expected value. The expected values come close to SDs, resulting in CoVs only fluctuating in a small range of [0.74 1.13]. Since a stronger correlation $r$ signifies the infectiousness of the epidemic model and tends to bring higher expected losses, the common cyber risk across TGs also increases. In Case Group 2, the incentive of investing on cyber-physical enhancement can be observed from Table V that expected losses are reduced substantially and reduction of SDs occurs to a lesser extent, with CoVs lying in [0.88 1.33].

TABLE IV Case Group 1: Expected Values (M$), Standard Deviations (M$) and Coefficients of Variation of Monetary Loss in the TGs

| $r = 0$ | TG1 | TG 2 | TG3 | TG4 | TG5 |
|---|---|---|---|---|---|
| $E[\mathcal{L}_q]$ | 4.42 | 7.10 | 2.76 | 3.49 | 3.92 |
| SD | 4.83 | 5.97 | 2.88 | 3.69 | 3.29 |
| CoV | 1.09 | 0.84 | 1.05 | 1.06 | 0.84 |
| $r = 0.5$ | TG1 | TG2 | TG3 | TG4 | TG5 |
| $E[\mathcal{L}_q]$ | 4.60 | 9.82 | 3.66 | 3.79 | 4.02 |
| SD | 5.04 | 10.5 | 3.51 | 4.29 | 2.99 |
| CoV | 1.10 | 1.07 | 0.96 | 1.13 | 0.74 |
| $r = 1$ | TG1 | TG2 | TG3 | TG4 | TG5 |
| $E[\mathcal{L}_q]$ | 7.45 | 12.7 | 3.97 | 4.74 | 5.28 |
| SD | 5.89 | 11.2 | 3.34 | 4.11 | 4.67 |
| CoV | 0.79 | 0.88 | 0.84 | 0.87 | 0.88 |

TABLE V Case Group 2: Expected Values (M$), Standard Deviations (M$) and Coefficients of Variation of Monetary Loss in the TGs

| $r = 0$ | TG1 | TG2 | TG3 | TG4 | TG5 |
|---|---|---|---|---|---|
| $E[\mathcal{L}_q]$ | 2.48 | 3.96 | 1.94 | 1.53 | 1.82 |
| SD | 3.30 | 3.83 | 2.04 | 1.91 | 1.91 |
| CoV | 1.33 | 0.97 | 1.05 | 1.25 | 1.05 |
| $r = 0.5$ | TG1 | TG2 | TG3 | TG4 | TG5 |
| $E[\mathcal{L}_q]$ | 2.79 | 5.73 | 2.60 | 1.89 | 2.69 |
| SD | 3.72 | 6.92 | 2.30 | 2.46 | 2.61 |
| CoV | 1.33 | 1.21 | 0.88 | 1.30 | 0.97 |
| $r = 1$ | TG1 | TG2 | TG3 | TG4 | TG5 |
| $E[\mathcal{L}_q]$ | 5.02 | 7.01 | 3.70 | 2.19 | 3.06 |
| SD | 4.95 | 6.99 | 3.51 | 2.38 | 3.46 |
| CoV | 0.99 | 1.00 | 0.95 | 1.09 | 1.13 |

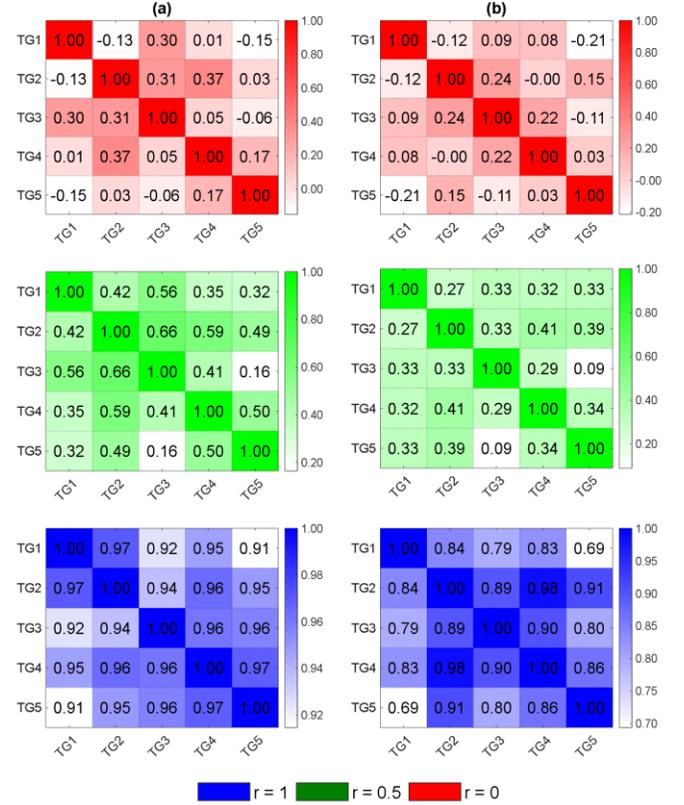

Fig. 7. SoI of the loss profile in the TGs in (a) Case Group 1 (b) Case Group 2.

In Fig. 7, the sampled SoI among the TGs are demonstrated in the Pearson correlation matrix. The correlation is symmetric and correlation between each of the two TGs can be observed in the off-diagonal entries. Fig. 8(a) depicts the correlation matrix of the Case Group 1. When $r = 0$, the SoI across the TGs are close to 0 with higher correlations between the neighboring TGs in the same areas. The correlations range around 0.45 as $r$ increases to 0.5. When $r = 1$, the correlations across all TGs are above 0.9. The correlation matrix in Case Group 2 is as shown in Fig. 8(b). Due to reduced load losses, the correlations are in general weakened between the same pair of TGs in Case Group 1.

Insurance premiums are designed to prepare TGs for catastrophic losses induced by probable cyberattack events. For interconnected TGs, mutual insurance accounting for respective marginal loss statistics would be a sensible option. The premium with a high-risk loading offers solid indemnity, which may however be less financially appealing to potential participants. An ideal premium design should be meticulously formulated to avoid excessive financial burdens while providing sufficient loss indemnities for the insured parties. The highly infectious nature of the cyber epidemic model dictates a heavily skewed tail risk.

To validate the design of the proposed cyber-insurance principle, herein (a) TCE premium $\pi_1$, (b) Coalitional premium $\pi_2$, and (c) Shapley premium $\pi_3$ of this study are compared at various degrees of correlation of the TGs. The TCE Premium is the most conservative design predominantly responsive to the tail risk, providing great redundancy at the cost of high-risk loading. On the contrary, the Coalitional Premium is the most affordable package by excluding extreme high-loss events with low probabilities. The Shapley Premium is cooperative and tailored to add further coverage against the tail risk, striking a balance between the affordability and loss coverage.

To gauge the relative premium burden against the expected risk, RLC is defined as follows:

$$\rho(\mathcal{L}_q) = \pi(\mathcal{L}_q)/E[\mathcal{L}_q] - 1 \qquad (12)$$

where $\rho(\mathcal{L}_q)$ should be generally positive to gather sufficiency budget for loss coverage. While positive RLC is preferable against the unexpected extreme risk, excessively high RLC would discourage the TGs from insurance participation.

In [17], the indemnities of $\pi_1(\mathcal{L}_q)$ are not clearly specified since the original design is tailored to a third-party insurer. In this paper, all premium designs are assumed to be mutual insurance. All participating entities are both insurers and insureds. For the sake of brevity, the indemnities of $\pi_1(\mathcal{L}_q)$ are proportionally allocated based on $\Gamma_q^\psi(\pi_2)$:

$$\Gamma_q^\psi(\pi_1) = \sum_q \pi_1(\mathcal{L}_q) * \frac{\Gamma_q^\psi(\pi_2)}{\sum_q \Gamma_q^\psi(\pi_2)} \qquad (13)$$

In Tables VI and VII, $\pi_1$, $\pi_2$, and $\pi_3$ are evaluated based on the loss statistics extracted from the two case groups with heavy tail risks. Characteristics of each design will be further elaborated numerically as follows.

The premiums of Case Group 1 are shown in Table VI. In each TG, $\pi_1$, $\pi_2$, and $\pi_3$ are positively correlated with the strength of correlation $r$. $\pi_1$ has the most conservative payment schedule and can be financially burdensome. $\pi_1$ may penalize the participants with heavy risk loading when extreme catastrophic events do not happen. Cost-effectiveness of $\pi_1$ is unacceptably low because the maximum of $\rho_1$ exceeds 3. On the flip side, $\pi_2$ is an entry-level premium design devised to be the most affordable and evenly distributed package across the TGs. $\pi_2$ offers small indemnities and the premiums collected from the TGs.

TABLE VI Actuarial Insurance Premiums (M$) in Case Group 1

| | TG1 | TG2 | TG3 | TG4 | TG5 |
|---|---|---|---|---|---|
| $r = 0$ | **TG1** | **TG2** | **TG3** | **TG4** | **TG5** |
| $\pi_1$ | 19.0 | 20.3 | 11.3 | 12.6 | 12.6 |
| $\Gamma_q^\psi(\pi_1)$ | 15.3 | 18.8 | 13.1 | 14.1 | 14.6 |
| $\rho_1$ | 3.30 | 1.87 | 3.11 | 2.60 | 2.22 |
| $\pi_2$ | 4.93 | 6.51 | 4.35 | 4.78 | 5.07 |
| $\Gamma_q^\psi(\pi_2)$ | 8.74 | 10.7 | 7.49 | 8.04 | 8.36 |
| $\rho_2$ | 0.12 | -0.08 | 0.58 | 0.37 | 0.29 |
| $\pi_3$ | 4.06 | 7.97 | 3.25 | 4.92 | 5.91 |
| $\Gamma_q^\psi(\pi_3)$ | 17.0 | 18.1 | 10.1 | 11.2 | 11.3 |
| $\rho_3$ | -0.08 | 0.12 | 0.18 | 0.41 | 0.51 |
| $r = 0.5$ | **TG1** | **TG2** | **TG3** | **TG4** | **TG5** |
| $\pi_1$ | 19.9 | 40.7 | 12.3 | 13.5 | 12.8 |
| $\Gamma_q^\psi(\pi_1)$ | 19.0 | 26.5 | 17.7 | 17.8 | 18.2 |
| $\rho_1$ | 3.32 | 3.14 | 2.37 | 2.57 | 2.19 |
| $\pi_2$ | 5.33 | 7.37 | 5.28 | 5.26 | 5.48 |
| $\Gamma_q^\psi(\pi_2)$ | 9.93 | 13.8 | 9.22 | 9.31 | 9.49 |
| $\rho_2$ | 0.16 | -0.25 | 0.44 | 0.39 | 0.36 |
| $\pi_3$ | 4.39 | 8.26 | 6.47 | 5.99 | 6.81 |
| $\Gamma_q^\psi(\pi_3)$ | 11.6 | 23.7 | 7.17 | 7.85 | 7.46 |
| $\rho_3$ | -0.05 | -0.16 | 0.77 | 0.58 | 0.69 |
| $r = 1$ | **TG1** | **TG2** | **TG3** | **TG4** | **TG5** |
| $\pi_1$ | 20.5 | 40.8 | 12.5 | 13.9 | 17.9 |
| $\Gamma_q^\psi(\pi_1)$ | 21.8 | 27.9 | 17.8 | 18.7 | 19.3 |
| $\rho_1$ | 1.75 | 2.20 | 2.14 | 1.94 | 2.39 |
| $\pi_2$ | 7.55 | 9.39 | 6.03 | 6.40 | 6.44 |
| $\Gamma_q^\psi(\pi_2)$ | 14.1 | 18.1 | 11.5 | 12.1 | 12.5 |
| $\rho_2$ | 0.01 | -0.26 | 0.52 | 0.35 | 0.22 |
| $\pi_3$ | 9.85 | 10.9 | 7.21 | 8.28 | 6.86 |
| $\Gamma_q^\psi(\pi_3)$ | 16.2 | 32.2 | 9.84 | 11.0 | 14.2 |
| $\rho_3$ | 0.32 | -0.15 | 0.81 | 0.75 | 0.30 |

TABLE VII Actuarial Insurance Premiums (M$) in Case Group 2

| | TG1 | TG2 | TG3 | TG4 | TG5 |
|---|---|---|---|---|---|
| $r = 0$ | **TG1** | **TG2** | **TG3** | **TG4** | **TG5** |
| $\pi_1$ | 12.1 | 12.2 | 7.51 | 7.49 | 7.60 |
| $\Gamma_q^\psi(\pi_1)$ | 9.47 | 10.5 | 9.11 | 8.84 | 9.03 |
| $\rho_1$ | 3.90 | 2.08 | 2.87 | 3.91 | 3.18 |
| $\pi_2$ | 2.80 | 3.77 | 2.77 | 2.43 | 2.67 |
| $\Gamma_q^\psi(\pi_2)$ | 4.91 | 5.42 | 4.72 | 4.58 | 4.68 |
| $\rho_2$ | 0.13 | -0.05 | 0.43 | 0.60 | 0.47 |
| $\pi_3$ | 2.21 | 4.51 | 2.78 | 0.87 | 1.90 |
| $\Gamma_q^\psi(\pi_3)$ | 7.72 | 7.75 | 4.78 | 4.77 | 4.84 |
| $\rho_3$ | -0.11 | 0.14 | 0.43 | -0.43 | 0.04 |
| $r = 0.5$ | **TG1** | **TG2** | **TG3** | **TG4** | **TG5** |
| $\pi_1$ | 13.7 | 26.4 | 8.00 | 9.19 | 8.19 |
| $\Gamma_q^\psi(\pi_1)$ | 12.5 | 17.1 | 12.2 | 11.1 | 12.4 |
| $\rho_1$ | 3.89 | 3.60 | 2.07 | 3.87 | 2.04 |
| $\pi_2$ | 3.22 | 4.37 | 3.52 | 2.89 | 3.57 |
| $\Gamma_q^\psi(\pi_2)$ | 6.02 | 8.23 | 5.88 | 5.34 | 5.95 |
| $\rho_2$ | 0.15 | -0.24 | 0.35 | 0.53 | 0.32 |
| $\pi_3$ | 2.27 | 4.69 | 5.30 | 1.32 | 2.42 |
| $\Gamma_q^\psi(\pi_3)$ | 7.41 | 14.3 | 4.34 | 4.98 | 4.44 |
| $\rho_3$ | -0.19 | -0.18 | 1.04 | -0.29 | -0.10 |
| $r = 1$ | **TG1** | **TG2** | **TG3** | **TG4** | **TG5** |
| $\pi_1$ | 17.3 | 26.7 | 11.8 | 9.95 | 14.2 |
| $\Gamma_q^\psi(\pi_1)$ | 17.2 | 20.0 | 15.3 | 13.1 | 14.4 |
| $\rho_1$ | 2.45 | 2.80 | 2.19 | 3.54 | 3.64 |
| $\pi_2$ | 5.18 | 5.84 | 4.72 | 3.78 | 4.08 |
| $\Gamma_q^\psi(\pi_2)$ | 9.01 | 10.5 | 8.02 | 6.89 | 7.54 |
| $\rho_2$ | 0.03 | -0.17 | 0.28 | 0.72 | 0.33 |
| $\pi_3$ | 6.24 | 6.55 | 6.10 | 1.80 | 3.11 |
| $\Gamma_q^\psi(\pi_3)$ | 10.6 | 16.3 | 7.20 | 6.07 | 8.65 |
| $\rho_3$ | 0.24 | -0.07 | 0.65 | -0.18 | 0.02 |

TABLE VIII Insolvency Probability (%) of Actuarial Insurance Premiums in Case Groups 1, 2

| Case Group 1 | | | | | |
|---|---|---|---|---|---|
| $r = 0$ | **TG1** | **TG2** | **TG3** | **TG4** | **TG5** |
| $\Phi(\pi_1)$ | 7.50 | 7.50 | 0 | 0 | 0 |
| $\Phi(\pi_2)$ | 10.0 | 17.5 | 7.50 | 15.0 | 10.0 |
| $\Phi(\pi_3)$ | 7.50 | 7.50 | 5.00 | 7.50 | 5.00 |
| $r = 0.5$ | **TG1** | **TG2** | **TG3** | **TG4** | **TG5** |
| $\Phi(\pi_1)$ | 5.00 | 7.50 | 0 | 0 | 0 |
| $\Phi(\pi_2)$ | 7.50 | 27.5 | 12.5 | 17.5 | 5.00 |
| $\Phi(\pi_3)$ | 7.50 | 7.50 | 12.5 | 17.5 | 5.00 |
| $r = 1$ | **TG1** | **TG2** | **TG3** | **TG4** | **TG5** |
| $\Phi(\pi_1)$ | 0 | 12.5 | 0 | 0 | 0 |
| $\Phi(\pi_2)$ | 15.0 | 20.0 | 7.50 | 10.0 | 10.0 |
| $\Phi(\pi_3)$ | 15.0 | 12.5 | 10.0 | 15.0 | 10.0 |
| Case Group 2 | | | | | |
| $r = 0$ | **TG1** | **TG2** | **TG3** | **TG4** | **TG5** |
| $\Phi(\pi_1)$ | 7.50 | 10.0 | 0 | 0 | 2.50 |
| $\Phi(\pi_2)$ | 20.0 | 25.0 | 10.0 | 5.00 | 2.50 |
| $\Phi(\pi_3)$ | 7.50 | 20.0 | 10.0 | 5.00 | 2.50 |
| $r = 0.5$ | **TG1** | **TG2** | **TG3** | **TG4** | **TG5** |
| $\Phi(\pi_1)$ | 7.50 | 7.50 | 0 | 0 | 0 |
| $\Phi(\pi_2)$ | 10.0 | 27.5 | 15.0 | 7.50 | 20.0 |
| $\Phi(\pi_3)$ | 7.50 | 7.50 | 15.0 | 7.50 | 20.0 |
| $r = 1$ | **TG1** | **TG2** | **TG3** | **TG4** | **TG5** |
| $\Phi(\pi_1)$ | 2.50 | 7.50 | 0 | 0 | 2.50 |
| $\Phi(\pi_2)$ | 12.5 | 25.0 | 15.0 | 5.00 | 7.50 |
| $\Phi(\pi_3)$ | 10.0 | 7.50 | 15.0 | 5.00 | 7.50 |

from -0.26 to 0.58. By contrast, $\rho_3$ is dispersed in [-0.16 0.81], a typical range of risk loading. $\pi_3$ offers a wider margin in risk loading than $\pi_2$ to guarantee sufficient budget to cover individual risk.

In Table VII, risk loading in Case Group 2 generally increase due to the enhanced security measure that reduces tail risk profile. $\rho_1$ has a maximum close to 4 and could be too high to motivate entities to participate in. $\pi_2$ is evenly distributed against average risk, with $\rho_2$ lying in [-0.24 0.72]. $\pi_3$ renders ideal risk loading $\rho_3$ to rarely exceed 1. High capacity of indemnity and low risk loading make the proposed $\pi_3$ a potentially compelling insurance model in practice.

The probability of insolvency $\Phi(\pi)$ is another risk measure which quantifies the capability of the insurance to mitigate the insolvency. $\Phi(\pi)$ is defined as the probability that the loss is greater than the indemnity:

$$\Phi(\pi) = \Pr[\mathcal{L}_q > \Gamma_q^\psi(\pi)] \quad (14)$$

As shown in Table VIII, in Case Group 1, $\pi_1$ generally provides the best insolvency alleviation with lowest probabilities of insolvency. In fact, $\pi_1$ is such a conversative premium design against risk that the insolvency in some cases is 0. While $\pi_3$ leads to the insolvency being lower than $\pi_2$ and greater than $\pi_1$, $\pi_3$ has the affordability superior to $\pi_1$. In Case Group 2, when the cyber risk is significantly reduced, $\pi_3$ can restrain the insolvency to be about as low as that of $\pi_1$. Thus, $\pi_3$ offers an economical option with relatively sufficient insolvency mitigation.

## V. CONCLUDING REMARKS

In this paper, a mutual insurance premium principle is designed to fairly share cyber risks across the participating TGs and control the overall insolvency risk. This study is among the first endeavors to approach the cyber-insurance by estimating the insolvency. In the case studies, it is shown the smart monitoring and job thread assignment solutions can work standalone or together to boost the reliability of TGs. Reduced insolvency probability is offered by the proposed Shapley premium while remaining as affordable as the coalitional

$\rho_2$ of some TGs can be slightly negative with indemnities supplemented by other TGs. However, the worse risk of $\pi_2$ beyond expected losses could barely be covered. $\pi_3$ rewards TGs of relatively low risk loading with high indemnities. While $\pi_1$ provides higher indemnities than $\pi_3$, $\pi_3$ offers comparable affordability to the coalitional platform of $\pi_2$. The proposed $\pi_3$ substantially alleviates the insolvency hazard of $\pi_2$. $\rho_2$ spans premium. More challenges may occur when real-life variables are factored in. Since any two power system servers are to some extent connectable from each other, establishing the topology of cyber node connections could be complicated. Selecting weights to prioritize the crucial edges in the cyber node graph could be essential. There are also challenges on the actuarial end. First, accurate cyber risk estimation for specific systems would rely on long-term historical data set collection. How much risk loading a premium design reserved should be sufficient against tail risk is still left to further exploration. Second, the proposed Shapley insurance scheme is designed to achieve two goals: insolvency risk control and fair distribution of indemnity. Although these goals are achieved most of the time, there are exceptions especially when some participants are struck by unexpectedly high losses due to inadequate self-protection. This shall motivate future work in designing more insurance schemes to reflect self-protection level and thus incentivize cyber-security investment.